# A Recursive Method for Enumeration of Costas Arrays

Mojtaba Soltanalian and Petre Stoica

Dept. of Information Technology, Uppsala University, Uppsala, Sweden

*Abstract—In this paper, we propose a recursive method for finding Costas arrays that relies on a particular formation of Costas arrays from similar patterns of smaller size. By using such an idea, the proposed algorithm is able to dramatically reduce the computational burden (when compared to the exhaustive search), and at the same time, still can find all possible Costas arrays of given size. Similar to exhaustive search, the proposed method can be conveniently implemented in parallel computing. The efficiency of the method is discussed based on theoretical and numerical results.*

## 1. INTRODUCTION

The concept of Costas arrays has been studied in engineering and mathematics for around half century; however, many fundamental questions are not yet answered [1]. Costas arrays seem to suggest a challenge for our present methodology in discrete mathematics [1][2]. The problem is easy to understand but has been difficult to tackle. Due of such difficulties, researchers have been very interested in computer search for Costas arrays.

A Costas array is simply a set of n points lying on the squares of a n×n checkerboard, such that each row and column contains only one point, and all of the $\binom{n}{2}$ displacement vectors between each pair of dots are distinct. Costas arrays are mainly known as time-frequency patterns that optimize the performance of sonars and radars. They also have applications in data hiding and mobile radio [3][4]. The application of Costas arrays in sonars and radars can be seen more clearly by an alternative definition of Costas arrays:

**Defintion 1.** A permutation matrix ($P$) of order n is a Costas array if and only if for any pair of integers $(r,s) \neq (0,0)$, $|r| \leq n$, $|s| \leq n$, the correlation function of the elements of $P$ satisfies

$$c(r,s) = \sum_{i=1}^{n} \sum_{j=1}^{n} P_{ij} P_{(i+r)(j+s)} \leq 1.$$

In order to identify the location and speed of a target, sonars and radars emit probing pulses at certain frequencies. The time delay between emission and reception indicates the distance of the target from the active sensing device. At the same time, due to the Doppler effect, the frequency difference between the emitted and the received signal indicates the velocity of the target. From this point of view, Costas arrays can be considered as a perfect time-frequency coding map.

There exist several construction methods for Costas arrays of sizes close to a prime number [5]. As a result of such constructions, Costas arrays are known for infinite number of sizes. The exhaustive search of Costas arrays has been accomplished to find and enumerate the Costas arrays of sizes up to n=27 [6]. To the best of our knowledge, n=32 and 33 are the smallest sizes for which no Costas arrays are known [7].

A Costas array and its properties can be investigated with a difference triangle, which its $m^{th}$ row represent the difference of the values located at indices with a distance of $m$ in the equivalent permutation array [1]. In particular, satisfaction of the Costas array property is equivalent to no duplicate entries in the rows of the associated difference triangle. To use the difference triangles, an exhaustive search for Costas arrays will include the formation of the difference triangle for all permutation matrices, along with performing all required comparisons. As the size grows large, the exhaustive search would lead to an extremely large computational burden. On the other hand, the computer search results have shown a rapid reduction in the number of Costas arrays for $16 \leq n \leq 26$. The number of Costas arrays for n=25, n=26, and n=27 are 88, 56 and 204, respectively [7]. It is also shown that the number of Costas arrays ($C_n$) satisfies $\lim_{n \to \infty}(C_n/n!) = 0$ (see [8] for a proof).

Given such computational difficulties, in this work, we propose a method that dramatically reduces the computational burden for finding Costas arrays. Moroever, the method enjoys the parallel computing capability that usually comes along with exhaustive search.

The rest of this paper is organized as follows: Section 2 is dedicated to a review on exhaustive search and its complexity. While the general idea of the method is discussed in Section 3, Section 4 aims to contribute an efficient scheme for Costas property inspection. Moreover, the implementation issues of the method are discussed in Section 5. Section 6 investigates the efficiency of the method from theoretical and numerical points of view. Finally, Section 7 concludes the paper.

## 2. THE EXHAUSTIVE SEARCH

Exhaustive or brute-force search over all permutation matrices has been a simple and straightforward method for finding Costas arrays in the past decades. This method has been able to find and enumerate all Costas arrays of sizes up to 29 [13]. The exhaustive search has been also the only method that can find all Costas arrays of given size--- due to the fact that some Costas arrays are "sporadic" i.e. they cannot be explained by currently known algebraic construction methods.

The main disadvantage of exhaustive search for Costas arrays is indeed the "combinatorial explosion", i.e. as the order increases, we quickly face an extremely large computational expense. The computational complexity of the exhaustive search for Costas arrays is $O(n^3 n!)$ where $n!$ denotes the number of all permutation matrices of size $n$, and $O(n^3)$ denotes the computational complexity of inspecting Costas property in a given permutation matrix. To see why, we should look at the difference triangle: In order to investigate the Costas property, we need $\binom{r}{2}$ comparisons in the $r^{th}$ row of the difference triangle and as a result, (using Chu's theorem) the total number of needed comparisons would be

$$\sum_{r=1}^{n-1} \binom{r}{2} = \binom{n}{3}.$$

The Equivalence of Costas arrays which is defined on rotations and reflections provides us with the possibility of reducing the number of Costas property inspections by an approximate factor of $1/8$; however, using the method for proportionally larger sizes like $n = 32$ seems yet prohibitive.

## 3. A RECUSRIVE CONFIGURATION

In this section, we propose a fairly simple recursive method with proportionally small computational burden in comparison to exhaustive search. Particularly, we show that a Costas array of given size can be constructed from smaller arrays which satisfy both of Permutation and Costas properties. This fact helps us to use the computational heritage that is available from the search of Costas arrays of smaller sizes.

Let $\chi(A)$ denote the minimum number of points that should be removed so that a permutation matrix A satisfy the Costas property. Also, let us define

$$S^n_{\chi \leq k} = \{all\ n \times n\ permutation\ matrices\ A | \chi(A) \leq k\}$$

It is interesting to note that a Costas array of size (n+1) can be constructed from $S^n_{\chi \leq 1}$: a Costas array of size (n+1) is nothing but a Costas array of size n with a corner point added, or (n-1) points in $n \times n$ checkerboard that satisfy both Permutation and Costas properties, with two points added in the row and column of the omitted point in the associated $n \times n$ permutation matrix. Fig. 1 illustrates the two described configurations.

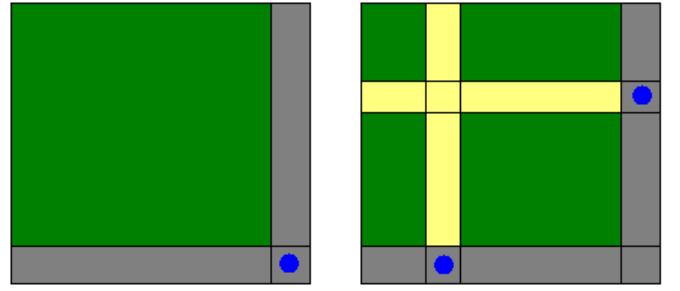

Figure 1. Configuration of points to construct a Costas array of size (n+1) from $S^n_{\chi \leq 1}$: The left configuration ($\chi = 0$), and the right configuration ($\chi = 1$).

Let $C(n, m)$ denote the number of $n \times n$ checkerboards with m points satisfying both Costas and Permutation properties. Also let $f(n, m)$ denote the number of candidates that our proposed method produces for Costas property inspection. Then it is straightforward to verify that

$$f(n, n) = C(n - 1, n - 1) + C(n - 1, n - 2)$$

Note that after inspecting the Costas property in all $f(n, n)$ candidates, we have obtained (and enumerated) all Costas arrays of size n.

Fortunately, beside the case of n points in the $n \times n$ checkerboard, other cases with smaller number of points on the $n \times n$ checkerboard can be constructed using a similar approach. Suppose we want to put $n - k$ points on an $n \times n$ checkerboard. The $n \times n$ checkerboard can be considered as an $(n - 1) \times (n - 1)$ checkerboard together with a region that the last row and last column make, which we call the new region. In the new region, zero, one or two points may exist. If the number of points in this region be equal to zero then all $n - k$ points must be on the $(n - 1) \times (n - 1)$ checkerboard. In the case of one point in the new region, $n - k - 1$ points must be on the $(n - 1) \times (n - 1)$ checkerboard and this makes $2k + 1$ possibility for placement of the one point in the new region. For two points, $n - k - 2$ points must be on the $(n - 1) \times (n - 1)$ checkerboard; this implies $(k + 1)^2$ possibility of placing the two points in the new region. The above discussion yields

$$\begin{aligned} f(n, n - k) &= C(n - 1, n - k) \\ &+ (2k + 1)C(n - 1, n - k - 1) \\ &+ (k + 1)^2 C(n - 1, n - k - 2). \end{aligned}$$

In the sequel, we denote the set of all of $n \times n$ checkerboards with m points that satisfy both Costas and Permutation properties with $\Phi^n_m$. The function $C$ represents the number of elements of $\Phi$, i.e. $C(n, m) = |\Phi^n_m|$. Based on the above discussion, $\Phi_{n-k}^n$ is constructable from $\Phi_{n-k}^{n-1}$, $\Phi_{n-k-1}^{n-1}$ and $\Phi_{n-k-2}^{n-1}$. Such an easy construction of $\Phi_{n-k}^n$ from $\Phi_{n-k}^{n-1}$, $\Phi_{n-k-1}^{n-1}$ and $\Phi_{n-k-2}^{n-1}$ is a very useful and fundamental tool that leads to a relatively efficient method for the search of Costas arrays.

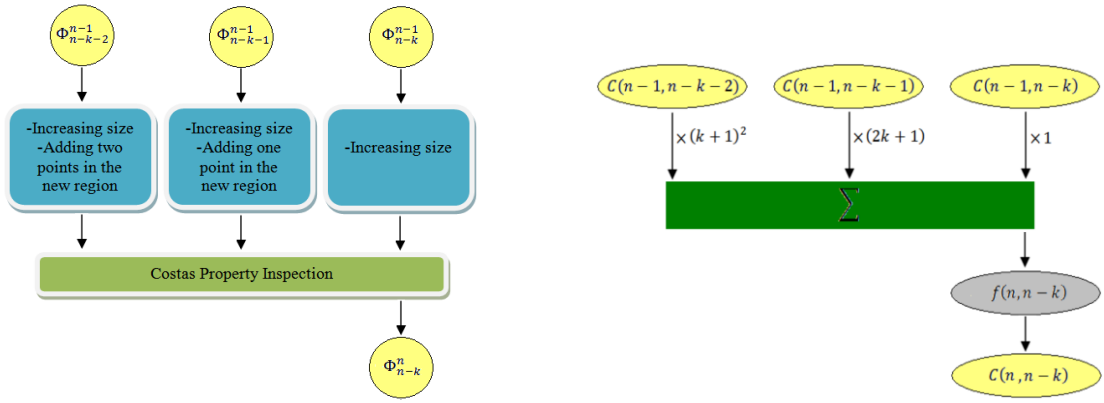

Figure 2. Construction of the elements of $\Phi$-Triangle and $C$-triangle from a triple of their upper elements.

**Defintion 2.** $C$**-Triangle** is a triangle of numbers whose $n^{th}$ row contains the values of the sequence $\{C(n,l)\}_{l=0}^n$.

Table 1 shows the $C$-triangle for $1 \leq n \leq 7$:

Table 1. the values of $C$-triangle for $1 \leq n \leq 7$.

| n/k | 0 | 1 | 2 | 3 | 4 | 5 | 6 | 7 |
|---|---|---|---|---|---|---|---|---|
| 1 | 1 | 1 | | | | | | |
| 2 | 1 | 4 | 2 | | | | | |
| 3 | 1 | 9 | 18 | 4 | | | | |
| 4 | 1 | 16 | 72 | 88 | 12 | | | |
| 5 | 1 | 25 | 200 | 568 | 720 | 40 | | |
| 6 | 1 | 36 | 450 | 2328 | 4412 | 2112 | 116 | |
| 7 | 1 | 49 | 882 | 7188 | 25592 | 32828 | 9844 | 200 |

Similar to the concept of $C$-triangle, a triangle of sets $\Phi$ could be defined:

**Definition 3.** $\Phi$**-Triangle** is a triangle of sets whose $n^{th}$ row consists of the set sequence $\{\Phi_l^n\}_{l=0}^n$.

The proposed construction also suggests to make every element of the $C$-triangle from a triple of its upper elements (the blank places can be assumed to be zero). In fact, every element of the $\Phi$-triangle is constructible from a triple of its upper elements since every element is a subset of specific structures which could be derived from the mentioned triple of sets. The result is clear, since the $C$-triangle just shows the cardinal of elements of the $\Phi$-triangle. Figure 2 illustrates the described idea herein.

To show an efficient recursive construction of the sets $\{\Phi_m^n\}$, suppose we already have all the sets $\{\Phi_{n-2r}^{n-r}\}$ and $\{\Phi_{n-2r+1}^{n-r}\}$ for $0 \leq r \leq \lfloor\frac{n}{2}\rfloor$. Suppose n is even; then $\Phi_0^{\frac{n}{2}}$ is constructible (directly or) from $\Phi_0^{\frac{n}{2}-1}$. If n was an odd number, $\Phi_1^{\lfloor\frac{n}{2}\rfloor}$ is constructable (directly or) from $\Phi_1^{\lfloor\frac{n}{2}\rfloor-1}$ and $\Phi_0^{\lfloor\frac{n}{2}\rfloor-1}$. Considering this as the first step, we should notice that like the previous results, $\Phi_{(n+1)-2r}^{(n+1)-r}$ is constructable from $\Phi_{n-2r+1}^{n-r}$, $\Phi_{n-2r}^{n-r}$ and $\Phi_{n-2r-1}^{n-r} = \Phi_{(n+1)-2(r+1)}^{(n+1)-(r+1)}$. This shows that the set sequence $\{\Phi_{(n+1)-2r}^{(n+1)-r}\}_{r=\lfloor\frac{n}{2}\rfloor}^0$ can be constructed step by step. The last element of the sequence represent the Costas arrays of size $(n+1)$. Taking the fact that $\Phi_{n-2r}^{n-r} = \Phi_{(n+1)-2(r+1)+1}^{(n+1)-(r+1)}$ into consideration, now we have all the sets $\Phi_{(n+1)-2r}^{(n+1)-r}$ and $\Phi_{(n+1)-2r+1}^{(n+1)-r}$ for $0 \leq r \leq \lfloor\frac{n+1}{2}\rfloor$, which completes a recursion. The discussed set sequences are depicted in Fig. 3 for $2 \leq n \leq 6$. All the mentioned sets should be available to obtain the Costas arrays of new size.

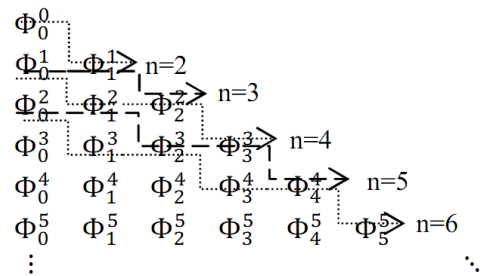

Figure 3. Required set sequences for $2 \leq n \leq 6$.

As one can observe, in each step, just one set in every column of $\Phi$-triangle should be saved. Let us denote the set in $k^{th}$ column by $\Phi[k]$. For even values of n, we should update $\Phi[k]$ s for even values of k. Also, $\Phi[k]$ s for odd values of k should be updated for the cases in which n is odd. The overall algorithm at the $n^{th}$ step of the method can be summarized as follows:

0. If n is even:
    0.0. For $k = 2, 4, \cdots, n$
        0.0.0. Update $\Phi[k]$ s consecutively.
1. Otherwise (if n is odd):
    1.0. For $k = 1, 3, \cdots, n$
        1.0.0. Update $\Phi[k]$ s consecutively.

Finally, it is clear that the last updated set (i.e. $\Phi[n]$) contains the Costas arrays of size n.

## 4. COSTAS PROPERTY INSPECTION

In this section, we derive an efficient Costas property inspection based on the recursive approach proposed earlier.

We discuss a generalization of the traditional Costas property inspection for the case of less than n points on $n \times n$ checkerboard. We should note that even with the traditional difference triangle, the number of needed comparisons for small numbers (k) of points is proportionally small. For the sake of intuition, consider a configuration of k points on the checkerboard. For these k points, just $\binom{k}{2}$ elements of $\binom{n}{2}$ elements in difference triangle exist. These $\binom{k}{2}$ available differences can be divided to $n-1$ parts (or sets), say the sets $S_1, S_2, \cdots, S_{n-1}$ which represent the rows of difference triangle. Note that $\sum_{i=1}^{n-1} |S_i| = \binom{k}{2}$, and moreover the number of needed comparisons would be equal to $\rho = \sum_{i=1}^{n-1} \binom{|S_i|}{2}$. Therefore,

$$\rho = \sum_{i=1}^{n-1} \binom{|S_i|}{2} = \frac{1}{2}\left(\sum_{i=1}^{n-1} |S_i|^2 - \sum_{i=1}^{n-1} |S_i|\right)$$
$$= \frac{1}{2}\left(\sum_{i=1}^{n-1} |S_i|^2\right) - \frac{1}{2}\binom{k}{2}$$

In the following, we provide an intuitive view on the upper bounds on the number of comparisons. Authors in [9] have introduced some benchmarks to assess the goodness of a sparsity measure. The relevant proposed benchmarks confirm the efficiency of the following sparsity measure for our problem.

**Defintion 3. Sparsity Measure-** Suppose a non-negative and constant sum sequence $\{x_i\}_{i=1}^n$. The summation

$$\sum_{i=1}^n x_i^2$$

is a sparsity measure for $\{x_i\}_{i=1}^n$: The bigger the sum, the more sparse the sequence.

Considering the above sparsity measure, we conclude that $\rho$ will be bigger if the vector

$$\vec{S} = \begin{pmatrix} |S_1| \\ |S_2| \\ \vdots \\ |S_{n-1}| \end{pmatrix}$$

is sparser. Since the sum of $|S_i|$ s is constant, making some $S_i$ s "richer" and making the others "poor" in the sense of the number of their elements, makes the vector sparser. The values $|S_i|$ are not independent, however, it can be shown that $min \, ||\vec{S}||_0 = k - 1$. The following lemma discusses such a minimal solution, as well as the configuration which leads to such a result.

**Lemma 1.** The minimum possible value of $||\vec{S}||_0$ is $k - 1$ and it occurs when the k points possess the same difference in a consecutive manner.

*Proof:* Consider k points in the locations
$$(a_1, \cdot), (a_2, \cdot), \cdots, (a_k, \cdot)$$
where
$$a_1 < a_2 < \cdots < a_k.$$

Since all the values $a_k - a_1, a_k - a_2, \cdots, a_k - a_{k-1}$ are distinct, $||\vec{S}||_0 \geq k - 1$. Now suppose that the $k - 1$ points $(a_1, \cdot), (a_2, \cdot), \cdots, (a_{k-1}, \cdot)$ yield exactly $k - 2$ distinct difference values if and only if
$$a_2 - a_1 = a_3 - a_2 = \cdots = a_{k-1} - a_{k-2}.$$

By adding the $k^{th}$ point, in order to obtain $||\vec{S}||_0 = k - 1$, just one value from the set $\{a_k - a_1, a_k - a_2, \cdots, a_k - a_{k-1}\}$ should be different from previous $k - 2$ distinct difference values. That would be $a_k - a_1$ as it is larger than any other difference value. Therefore, we have $a_k - a_2 = a_{k-1} - a_1$ and its direct consequence which is $a_k - a_{k-1} = a_2 - a_1$. This completes the proof. ∎

**Corollary 1.** The maximum number of needed comparisons for inspecting the Costas property for k points is $\binom{k}{3}$.

*Proof:* Intuitively, the minimal $||\vec{S}||_0$ in the above yields an upper bound on $||S||_2$. A proof of Corollary 1 is provided in the Appendix. ∎

Since the main method is based on recursion, a more convenient method of Costas property inspection could be designed. First suppose the k points in the locations $(a_1, \cdot), (a_2, \cdot), \cdots, (a_k, \cdot)$. Note that every comparison corresponds to a 3-element subset of $\{a_1, a_2, \cdots, a_k\}$. To explain this, we should remember that every comparison occurs on a 4-element set, viz. $\psi = \{x_1, x_1 + d, x_2, x_2 + d\}$. Suppose $x_1 < x_2$; therefore $x_1 < x_1 + d < x_2 + d$ and as a result, every 3-element subset of $\{a_1, a_2, \cdots, a_k\}$, say $\{a_p, a_q, a_r\}$ where $a_p < a_q < a_r$, is able to construct $\psi$ if we consider $(x_1, x_1 + d, x_2 + d) = (a_p, a_q, a_r)$. This yields $x_2 = a_p + a_r - a_q$ and the comparison is needed only if $x_2 = a_p + a_r - a_q \in \{a_1, a_2, \cdots, a_k\}$.

Now, we are ready to discuss a recursive approach in Costas property inspection. In the last section, we stated that in the new region, zero, one or two points may exist. If the number of points in the new region was zero, there is no need for new comparisons. In the case of one point in the new region, depending on the location of the point (whether it is on the last row or column), the row number or the column number of points would be like $a_1, a_2, \cdots, a_{k-1}, k$. Since the points in the $(n-1) \times (n-1)$ checkerboard satisfy the Costas property, we just need comparisons that involve the new point. Therefore, we just need to choose two elements from the set $\{a_1, a_2, \cdots, a_{k-1}\}$. The maximum number of such comparisons is $\binom{k-1}{2}$. Now suppose we want to put two points in the new region. As discussed above, we just need comparisons that involve at least one of the new points. The maximum number of needed comparisons would be $2\binom{k-2}{2} + \binom{k-2}{1} = (k-2)^2$. Therefore, the maximum number of needed comparisons to obtain $\Phi_{n-k}^n$ by our proposed method would be

$$\tilde{f}_u(n, n-k) = (2k+1)\binom{n-k-1}{2}C(n-1, n-k-1) + (k+1)^2(n-k-2)^2 C(n-1, n-k-2)$$

This upper bound will be used to investigate the efficiency of the method in Section 6.

## 5. REMARKS ON IMPLEMENTATION

In this section, we will focus on some implementation issues, including the starting point, compatibility for parallel search of Costas arrays, as well as the issues regarding the recursion steps.

1. **The starting point.** Typically, the starting point of the algorithm is a set of Costas patterns of small size ($n$), say $n = 1$. The output of each step contains the Costas patterns required for the next step. Note that even though the method is proposed to construct the Costas arrays of different size consequently; there is a fairly simple way to construct the needed set sequence for a specific n directly. It is easy to see that a set of points has the Costas property iff it has no parallelograms or three equidistant points on a same line. We call either of these two patterns "Costas Violator Pattern (CVP)". Now, the key idea is that $\Phi_{k+1}^n$ is constructable from $\Phi_k^n$: for every element of $\Phi_k^n$, considering the places not in the same rows or columns of the previous points, we can mark the places that putting a new point in them will make a CVP with the previous points. All non-marked such places are eligible to put the new point in order to make an element of $\Phi_{k+1}^n$. Fig. 4 has shown an example of applying this idea.

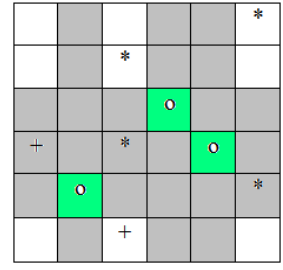

Figure 4. An example of applying the discussed idea to construct $\Phi_{k+1}^n$ from $\Phi_k^n$. Previous points, marked points to avoid a parallelogram pattern and points marked to avoid three equidistant points on a same line are denoted by o,+ and * respectively. Same row and same column locations with the previous points are also shown in gray. The figure shows that exactly 6 points can be used to construct members of $\Phi_4^6$ from the considered element of $\Phi_3^6$.

Using the discussed idea, by finding $\Phi_{n-2r}^{n-r}$ we can find $\Phi_{n-2r+1}^{n-r}$ and $\Phi_{n-2r+2}^{n-r}$ and as a result, having a knowledge of this triple, we can construct $\Phi_{n-2r+2}^{n-r+1} = \Phi_{n-2(r-1)}^{n-(r-1)}$. In this way, we are able to make the needed set sequences to find the Costas arrays of n+1.

2. **Compatibility for Parallel search.** One of the important issues is the compatibility of the method for parallel search of Costas arrays which is also an advantage of the exhaustive search as the set of n! permutation matrices can be partitioned to many parts in order to assign the work of Costas property inspection to several computing machines simultaneously. Clearly, a similar scenario can be used for the proposed recursive method, as for updating every Φ[k], candidates for Costas property inspection could be assigned to several machines. To optimize such assignment, since every candidate may need a different number of comparisons, we may assign a subset of candidates to a machine when it's going to an idle state. In fact, different schemes can be used to derive a parallel version of the basic algorithm and discussing them is beyond the scope of this paper.

3. **Recursion Steps.** It is also possible to change the recursion steps from what we have used in this paper. The basic method in this work is proposed to construct Costas arrays of different sizes consequently. Therefore, in the proposed method, every new Costas structure of size $n$ is supposed to be constructed from a Costas structure of size $(n-1)$. Surely, one can change this size-one step length and use larger sizes; for example, one can use the Costas structures of size $(n-2)$ to construct the Costas structures

Table 2. Comparison of real number of needed comparisons for the method $\tilde{f}_{rec.}(n)$, the discussed upper bound $\tilde{f}^u_{rec.}(n)$, the Exhaustive search $\tilde{f}_{ex.\ s.}(n)$ and the associated ratios for $3 \leq n \leq 10$.

| $n$ | $\tilde{f}_{rec.}(n)$ | $\tilde{f}^u_{rec.}(n)$ | $\tilde{f}_{ex.\ s.}(n)$ | $\frac{\tilde{f}_{rec.}(n)}{\tilde{f}_{ex.\ s.}(n)}$ | $\frac{\tilde{f}^u_{rec.}(n)}{\tilde{f}_{ex.\ s.}(n)}$ | $\frac{\tilde{f}_{rec.}(n)}{\tilde{f}_{rec.}(n-1)}$ | $\frac{\tilde{f}_{ex.\ s.}(n)}{\tilde{f}_{ex.\ s.}(n-1)} = \frac{n^2}{(n-3)}$ |
|---|---|---|---|---|---|---|---|
| 3 | 6 | 6 | 6 | 1.0000 | 1.0000 | - | - |
| 4 | 78 | 84 | 96 | 0.8125 | 0.8750 | 13.0000 | 16.0000 |
| 5 | 738 | 954 | 1200 | 0.6150 | 0.7950 | 9.4615 | 12.5000 |
| 6 | 6552 | 13864 | 14400 | 0.4550 | 0.9628 | 8.8780 | 12.0000 |
| 7 | 53784 | 88452 | 176400 | 0.3049 | 0.5014 | 8.2088 | 12.2500 |
| 8 | 419380 | 720032 | 2257920 | 0.1857 | 0.3189 | 7.7975 | 12.8000 |
| 9 | 3268280 | 6213020 | 30481920 | 0.1072 | 0.2038 | 7.7931 | 13.5000 |
| 10 | 25280816 | 53529728 | 435456000 | 0.0581 | 0.1229 | 7.7400 | 14.2857 |

of size (n). In such cases, a new (but similar) study of the problem can be accomplished.

## 6. REMARKS ON EFFICIENCY

In the following, the efficiency of our method is compared to the exhaustive search in both theoretical and numerical point of views. We use the number of needed comparisons as a benchmark for such judgement.

### 6.1. Efficiency- Theoretical Discussion

Typically, the number of comparisons needed in Exhaustive search is given by

$$\tilde{f}_{ex.\ s.}(n) = \binom{n}{3} n!$$

To obtain the number of comparisons needed in our proposed method, we use a summation of the number of needed comparisons for $\Phi[k]$ s' update process, which is discussed in Section 3. The upper bound of the number of needed comparisons would be

$$\tilde{f}^u_{rec.}(n) = \sum_{r=0}^{\left\lfloor \frac{n}{2} \right\rfloor} \tilde{f}_u(n-r, n-2r)$$

Table (2) shows a comparison of this upper bound for the proposed method $\tilde{f}^u_{rec.}(n)$, real number of needed comparisons for the method $\tilde{f}_{rec.}(n)$ and the Exhaustive search $\tilde{f}_{ex.\ s.}(n)$ for $3 \leq n \leq 10$. We should note that, $\tilde{f}_u(n,k)$ is a function of the values of $C$-triangle. The two following lemmas show two kinds of restrictions for values of $C$-triangle. The first one is a useful upper bound for first indices of the rows, whereas the second one aims to investigate an upper bound for values of $C$-triangle around the final indices of each row.

**Lemma 2.** For the elements of $C$-triangle, the following inequality holds:

$$C(n,k) \leq k! \binom{n}{k}^2$$

*Proof:* See the appendix for the proof. ∎

**Lemma 3.** For the elements of $C$-triangle, the following inequality holds:

$$C(n, n-k) < (n-k+1)C(n, n-k+1) + 4^k\, S^n_{\chi=k}$$

where
$$S^n_{\chi=k} = \{all\ n \times n\ permutations\ A | \chi(A) = k\}.$$

*Proof:* First of all, $\Phi^n_{n-k}$ can be constructed form $S^n_{\chi \leq k}$. Elements of $S^n_{\chi \leq k}$ can be transformed to elements of $\Phi^n_{n-k+1}$ by removing $(k-1)$ points, or they belong to $S^n_{\chi=k}$. Therefore, $\Phi^n_{n-k}$ can be constructed from $\Phi^n_{n-k+1}$ and $S^n_{\chi=k}$. The maximum number of new constructable elements of $\Phi^n_{n-k}$ from an element of $\Phi^n_{n-k+1}$ is $(n-k+1)$. On the other hand, If we need to remove at least k points from the elements of $S^n_{\chi=k}$, this means each element of $S^n_{\chi=k}$ has k "independent" CVPs. Each independent CVP suggest at most 4 points for removal, and as a result the upper bound on the number of new elements of $\Phi^n_{n-k}$ constructed from an element of $S^n_{\chi=k}$ would be $4^k$. These facts conclude the inequality. ∎

It is interesting to notice that since every CVP contains at least 3 points, the maximum number of independent CVPs would be $\left\lfloor \frac{n}{3} \right\rfloor$. Therefore, as we discussed before, there is no permutation matrix for $\chi > \frac{n}{3}$. This implies $\left| S^n_{\chi=k} \right| = 0$ for $k > \frac{n}{3}$.

**Corollary 2.** $\lim_{n \to \infty} (C(n, n-k)/(n+k)!) = 0$.

*Proof:* First of all, we know that

$$\lim_{n\to\infty}\frac{C(n,n)}{n!}=0.$$

Now suppose that $\lim_{n\to\infty}\frac{C(n,n-(k-1))}{(n+(k-1))!}=0$. Using Lemma 3, and employing the fact that $S^n_{\chi=k}<n!$, we can write

$$0<C(n,n-k)<(n-k+1)C(n,n-k+1)+4^k n!$$

or equivalently,

$$0<\frac{C(n,n-k)}{(n+k)!}<\left(\frac{n-k+1}{n+k}\right)\frac{C(n,n-k+1)}{(n+k-1)!}+\frac{4^k n!}{(n+k)!}$$

The above inequality yields

$$\lim_{n\to\infty}\frac{C(n,n-k)}{(n+k)!}=0$$

which completes the proof. ∎

We want to show that the amount of $\tilde{f}_{rec.}(n)$ is negligible in comparison to $\tilde{f}_{ex.\ s.}(n)$ when $n\to\infty$. We restate this as

$$\lim_{n\to\infty}\frac{\tilde{f}_{rec.}(n)}{\tilde{f}_{ex.\ s.}(n)}=0.$$

To show this, we will show a more general phenomenon. Since $\tilde{f}_{rec.}(n)\le\tilde{f}^u_{rec.}(n)$, it is sufficient to show that

$$\lim_{n\to\infty}\frac{\tilde{f}^u_{rec.}(n)}{\tilde{f}_{ex.\ s.}(n)}=\lim_{n\to\infty}\frac{\tilde{f}^u_{rec.}(n)}{n^3 n!}=0.$$

As discussed in the beginning of the section:

$$\tilde{f}^u_{rec.}(n)=\sum_{r=0}^{\lfloor\frac{n}{2}\rfloor}\tilde{f}_u(n-r,n-2r)$$

$$=\sum_{r=0}^{\lfloor\frac{n}{2}\rfloor}(2r+1)\binom{n-2r-1}{2}C(n-r-1,n-2r-1)$$

$$+\sum_{r=0}^{\lfloor\frac{n}{2}\rfloor}(r+1)^2(n-2r-2)^2 C(n-r-1,n-2r-2).$$

Using Corollary 2, we have

$$\lim_{n\to\infty}\frac{C(n-r-1,n-2r-1)}{(n-1)!}=0.$$

Therefore, for every value of $0\le r\le\lfloor\frac{n}{2}\rfloor$:

$$\lim_{n\to\infty}\frac{(2r+1)\binom{n-2r-1}{2}C(n-r-1,n-2r-1)}{n(n!)}=0$$

and as a result

$$\tilde{f}^{u1,\infty}_{rec.}(n)=\lim_{n\to\infty}\frac{\sum_{r=0}^{\lfloor\frac{n}{2}\rfloor}(2r+1)\binom{n-2r-1}{2}C(n-r-1,n-2r-1)}{n^3(n!)}=0.$$

Showing that the limitation for the second term of $\tilde{f}^u_{rec.}(n)$, (say $\tilde{f}^{u2}_{rec.}(n)$) approaches zero appears to be more challenging.

**6.2. Efficiency: Numerical Discussion**

In this subsection we aim to discuss the efficiency according to the numerical results in Table 2.

One can observe that the function $\frac{\tilde{f}_{rec.}(n)}{\tilde{f}_{ex.\ s.}(n)}$ is decreasing. Interestingly, the decrement rate of the values, defined as $a(n)=\left(\frac{\tilde{f}_{rec.}(n-1)}{\tilde{f}_{ex.\ s.}(n-1)}/\frac{\tilde{f}_{rec.}(n)}{\tilde{f}_{ex.\ s.}(n)}\right)$, is very well-fitted to the line $\tilde{a}(n)=0.8218+0.1024n$. Using extrapolations, we can estimate a computational burden reduce factor of $2.67\times 10^{-12}$ and $6.35\times 10^{-13}$ for interesting cases of n=32 and n=33 respectively, which are the smallest sizes that no Costas arrays are known for.

Variations of growth of both $\tilde{f}_{rec.}(n)$ and $\tilde{f}_{ex.\ s.}(n)$ are also interesting. We can observe that

$$\frac{\tilde{f}_{rec.}(n)}{\tilde{f}_{rec.}(n-1)}<\frac{\tilde{f}_{ex.\ s.}(n)}{\tilde{f}_{ex.\ s.}(n-1)}$$

for all values $4\le n\le 10$, mentioned in the table. It is interesting to note that, in contrast to the growth rate of $\tilde{f}_{ex.\ s.}(n)$, which is an increasing function of n for $n>6$, the growth rate of $\tilde{f}_{rec.}(n)$ is decreasing.

## 7. CONCLUSION

A recursive method of finding Costas arrays is proposed and some of its implementation issues are discussed. Theoretical and numerical results appear to confirm the efficiency of the method in comparison to the exhaustive search.

## APPENDIX

*Direct proof of Corollary 1.* As discussed in Section 4, every comparison corresponds to a 3-element subset of $\{a_1,a_2,\cdots,a_k\}$, say $\{a_p,a_q,a_r\}$ where $a_p<a_q<a_r$. We discussed that the comparison is needed only if $x_2=a_p+a_r-a_q\in\{a_1,a_2,\cdots,a_k\}$. Also, the structure that

$$a_2-a_1=a_3-a_2=\cdots=a_k-a_{k-1}$$

satisfies this constraint for every distinct $p, r, q \in \{1, 2, \cdots, n\}$ and the proof is complete.

*Proof of Lemma 2.* In order to put $k$ points in $n \times n$ checkerboard, a $k \times k$ sub-checkerboard should be selected. Number of such sub-checkerboards is $\binom{n}{k}^2$. Although, the $k$ points make $k!$ permutations in every chosen sub-checkerboards. Therefore, the number of all $n \times n$ checkerboards that contain $k$ points and satisfy the Permutation property would be $k!\binom{n}{k}^2$ and obviously

$$C(n, k) \leq k!\binom{n}{k}^2.$$